\documentclass[apjl]{emulateapj}

\usepackage{graphicx}
\usepackage{subfigure}
\usepackage{amsmath} 

\usepackage[colorlinks=true,citecolor=blue,urlcolor=magenta,breaklinks]{hyperref}
\usepackage{natbib}
\usepackage[T1]{fontenc}
\usepackage{newtxtext,newtxmath}
\begin{document}


%
\title{The Sterile-Active Neutrino Flavor Model: \\
the Imprint of Dark Matter on the Electron Neutrino Spectra}
\author{Il\'idio Lopes}
\email[]{ilidio.lopes@tecnico.ulisboa.pt}
\affiliation{ Centro de Astrof\'{\i}sica e Gravita\c c\~ao -- CENTRA,
Departamento de F\'{\i}sica, Instituto Superior T\'ecnico -- IST,\\ 
Universidade de Lisboa -- UL, Av. Rovisco Pais 1, 1049-001 Lisboa, Portugal \\}
\affiliation{Institut d'Astrophysique de Paris, UMR 7095 CNRS, \\  Universit\'e Pierre et Marie Curie, 98 bis Boulevard Arago, Paris F-75014, France} 

%
%
%

\begin{abstract}
Contact interactions between sterile neutrinos and dark matter particles in a hidden sector have been suggested as a good solution to simultaneously resolve the dark matter problem and anomalies in neutrino experiments.
In this non-standard particle physics model, sterile and active neutrinos change their   through vacuum oscillations and matter (or Mikheyev--Smirnov--Wolfenstein) oscillations, in which the latter  mechanism  of flavor oscillation depends strongly on the concentration of dark matter in the Sun's core. We found that a large concentration of dark matter in the Sun's interior changes substantially the shape of ${\rm ^8B}$ and $\rm hep$ electron neutrino spectra, but has an insignificant impact on the other neutrino spectra (i.e., $\rm pp$, $\rm pep$, $\rm ^7Be$ and $\rm ^{15}O$, $\rm ^{13}N$ and  $\rm ^{17}F$).  
The strength of the interaction of the dark matter particles with neutrinos depends on an effective coupling constant,
 $G_\chi$, which is an analog  of the Fermi constant for the hidden sector.  By using the latest $\rm ^8B$ solar neutrino flux,  we found that $G_\chi$ must be smaller than $\rm  0.5\times 10^9$ $G_{\rm F}$ for this particle physics model to be in agreement with the data. 
\end{abstract}


\maketitle


\section{Introduction\label{sec-intro}}

\smallskip
Never has new particle been proposed to solve  so many problems in theoretical physics as the sterile neutrino. This particle is in the race to become the best candidate to resolve simultaneously some of the most fundamental problems in particle physics and cosmology~\citep{2017PhR...711....1A}. The sterile neutrino occurs naturally in  well-motivated  extensions of the standard model of particle physics. Its existence emerges naturally in one of the simplest mechanisms to explain the origin of the light  masses of active neutrinos, the so-called {\it see-saw} mechanism
~\citep[e.g.,][]{2013arXiv1306.4669G}. 
More significantly, this particle improves the agreement between theory and observations in the two complementary fields
of research: experimental particle  physics and observational cosmology.  Its potential  success in resolving such problems should not be underestimated.
Two of these highlights are worth mentioning.

\smallskip

First, the sterile neutrino can explain the origin of the 
anomalies observed in several neutrino experiments~\citep{2016JPhG...43c3001G}.  
Over the years there have been several discrepancies 
from various short-baseline neutrino experiments~\citep{2001PhRvD..64k2007A,2010PhRvL.105r1801A,2011PhRvD..83g3006M,2012AIPC.1441..458Z} that point to the possibility that the active neutrino flavor oscillation model is incomplete. This weakness of the neutrino oscillation model has led many studies to incorporate additional single states into the same framework~\citep[e.g.,][]{2007PhRvD..76i3005M,2012PhRvD..85a1302B}.  Even the simplest sterile neutrino model,
the 3+1 flavor oscillation model made of 3 active neutrinos
$(\nu_e,\nu_{\tau},\nu_{\mu})$ and a single sterile neutrino $\nu_s$, i.e.,
$(\nu_e,\nu_{\tau},\nu_{\mu},\nu_s)$,  parametrized by only two additional parameters --  a mass square splitting  and a sterile-active mixing angle -- seems to be a good solution~\citep{2011PhRvD..83k3013P}. This popular model is able to resolve  resolve the anomalies from the Liquid Scintillator Neutrino Detector~\citep{2001PhRvD..64k2007A} and MiniBooNE~\citep{2010PhRvL.105r1801A}, and also be consistent with the data coming from solar and atmospheric neutrino experiments, by choosing a mass-squared splitting of the order of $\rm 1 {\rm eV^2}$ and a mixing angle  of the order of $10^{-1}$~\citep{2011PhRvD..83g3006M}. Similar parameters are also seemingly required to reconcile the flux deficit of 
reactor antineutrinos~\citep{2011PhRvD..83g3006M}. This  interpretation has recently been challenged by ~\citet{2017PhRvL.118y1801A}: 
it seems that although its effect is less pronounced,   
the anomaly persists~\citep{2017arXiv170707728H}.
   
\smallskip
Second, sterile neutrinos were proposed  to resolve some of the well-known disagreements  between numerical simulations and observations in the formation of the large-scale structure in the universe~\citep{2014PhRvL.112c1803D}. The standard collisionless cold dark matter (CCDM) model predicts the formation of too much structure  formation at certain cosmological scales, in direct conflict with astronomical observations~\citep{2017FrPhy..12l1201Y}. These differences have been identified in three cases: (i) the {\it galaxy cusp core problem}, the disagreement between the cuspy density profiles predicted by numerical simulations of  CCDM and the core profiles found in dwarf galaxies~\citep[e.g.,][]{1994Natur.370..629M};
(ii) the {\it the too-big-to-fail problem}, the fact that most large sub-haloes in CCDM simulations are too massive to host the satellites of the Milky Way~\citep{2011MNRAS.415L..40B}; (iii) the {\it missing satellites problem}, wherein the number of satellites found in simulations of the  Milky Way-sized halos
disagrees with observations by roughly a factor 10~\citep{1993MNRAS.264..201K}.

\smallskip
\smallskip
Although the first set of problems is well resolved by sterile neutrinos, which suggests that  these neutrino experimental anomalies can be explained by postulating a sterile neutrino with
the following parameters: mass difference $\rm \sim eV$ and a  mixing angle $\rm \sim 0.1\;(or\; 6^o)$, a detailed study has shown that these sets of parameters do not fix completely the problem found in  the formation  of structure~\citep{2011JCAP...09..034H,2014PhRvL.112c1802H}. 
To alleviate this difficulty, a new mechanism has been postulated, the existence of a hidden sector comprising dark matter particles and sterile neutrinos in which the interactions between them are mediated by a massive gauge boson. The term {\it hidden sector} reinforces the fact that these interactions happen outside the standard model of particle physics. The existence of gauge forces within the dark matter sector has been considered for a long time. Actually, many of these models have been tested and constrained using several cosmological and astrophysical observational data sets~\citep[e.g.,][]{2010ARNPS..60..405J}.
The possibility that these types of interactions can also occur with the neutrino sector seems a good one.  

\smallskip  
In this new class of cosmological models it is assumed that in the early universe there is an interaction between dark matter and sterile neutrinos, which, unlike typical cold dark matter candidates, keeps the dark matter kinetically coupled until a much later epoch. This change impacts the abundance of primordial elements during the Big Bang nucleosynthesis, modifies the cosmic microwave background (CMB), and alters the  formation of  large-scale structure, eliminating the conflict
between simulations and data~\citep{2016A&A...594A..13P}. Moreover, the observational galactic problems mentioned above, such as the creation of dark matter cusps in the core of galaxies and a surplus of galaxies in the Local Group 
are successfully resolved in these cosmological models, which, unlike the standard cosmological model, do not lead to an excess in the formation of structure~\citep{2007PhRvD..76j3515H,2014arXiv1411.1071C,2015JHEP...04..170B}. This better agreement between  observations and the theory is mainly due to the   oscillations between active and sterile neutrinos. 

\smallskip
As previously mentioned, sterile neutrinos are most likely  created in the primitive universe~\citep[e.g.,][]{1992PhLB..275..112B} or inside stars~\citep[e.g.,][]{2018EPJC...78..327L}, either as the final product of the annihilation of dark matter particles or as a result of neutrino flavor oscillations.
Indeed, these neutrinos appear to be the final state of annihilating dark matter in many explicit particle physics models previously considered in the literature~\citep[e.g.,][]{2017JHEP...05..102G}, and in other cases these are actually the dominant annihilation channel for dark matter~\citep[e.g.,][]{2014PhRvD..90g5021D}. In any case, these neutrinos must have a mass below $10^3$ TeV~\citep{2017PhRvD..95i5016B}, a mass range that easily encompasses the expected mass of many dark matter particle candidates~\citep{2015PhR...555....1B}.  
The most important aspect of the class of particle physics models that is being considered in this study  is the fact that neutrinos must interact at least in part  with the dark matter, hence opening a window to this hidden sector.

\smallskip
The propagation of neutrinos throughout space may be affected 
by hypothetical couplings to the cosmic medium, such as dark matter,
dark radiation, and dark energy. This  challenging question  has been addressed by many  authors~\citep{2006JCAP...12..013C,2018PhRvD..97j3004C,2018PhRvD..97f3529G}. Recently, \citet{2018JCAP...07..004C} found that the next generation of detectors such as the Neutrino Telescope~\citep[KM3NeT, e.g.,][]{2016JPhG...43h4001A}  and IceCube Neutrino Observatory 
~\citep[IceCube, e.g.,][]{2017PhRvD..95k2002A} have the necessary sensitivity
to probe many of the proposed non-standard  interactions 
of neutrinos. \citet{2018JCAP...07..004C} found these effects can be constrained to a level of a few per cent on the Earth's matter potential with coupling  mediated by the $\mu$-neutrinos. Moreover,  baseline experiments such as
Deep Underground Neutrino Experiment ~\citep[e.g.,][]{2013arXiv1307.7335L} may provide additional
complementary constraints on neutrino properties in the  dark sector.
Following the same strategy, these authors have shown that solar neutrino
data can be used to put constraints on the  3+1 neutrino model in
which neutrinos interact with dark matter by means of a 
Mikheyev--Smirnov--Wolfenstein
(MSW) mechanism~\citep{2017JCAP...07..021C}. They found that
such modifications affect $\rm ^8B$ and 
carbon--nitrogen--oxygen (CNO) neutrinos.

\smallskip
The Sun and sun-like stars have been used quite robustly to impose constraints on large number of dark matter models~\citep{2012RAA....12.1107T}. This is possible because the accumulation of dark matter inside the star changes locally the transport of the energy affecting its whole structure, especially the core. Indeed, \citet{2002MNRAS.331..361L} have shown that the presence of dark matter significantly modifies the local luminosity of the star's core by changing the sound speed and temperature profiles, leading to the possibility of
 placing putting stringent constraints
on such dark matter models using  observational astronomical data. This is done in the Sun's case by comparing results of stellar models with helioseismic observations and data
on  solar neutrino flux, and in the case of sun-like stars by comparing with  asteroseismic observations~\citep{2012RAA....12.1107T}.
In recent years a multitude of different dark matter particle candidates have been suggested, all of which are able to affect the structure of the star. Nevertheless, depending on the specific properties of the dark matter particles, they can produce more or less significant changes.
For instance, several authors have shown that self-interacting asymmetric dark matter can significantly enhance the accretion of dark matter by the star~\citep{2010PhRvD..82j3503C,2010PhRvL.105a1301F,2010PhRvD..82h3509T,2012ApJ...757..130L}. Alternatively,~\citet{2014ApJ...795..162L} suggest that an asymmetric dark matter particle with a  momentum-dependent scattering cross section could improve the agreement between the solar model and the helioseismic data, resolving  the solar composition problem~\citep{2016JCAP...11..007V}. 
In addition,  the Sun is no longer the only star we use to constrain the properties of dark matter. 
\citet{2011MNRAS.410..535C} show that the oscillations of sun-like stars can be used as diagnostics to constrain the properties of dark matter  further.  In a follow-up to this work, the same group have reported the first asteroseismic constraints for
asymmetric dark matter models~\citep{2013ApJ...765L..21C,2017PhRvD..95b3507M}.
Equally, many other dark matter models have been presented in the literature~\citep[e.g.,][]{2011PhRvL.107i1301K,2012PhRvD..85j3528L}.

\smallskip
The standard theory of neutrino flavor oscillations based on three massive neutrinos~\citep[e.g.,][]{2008PhR...460....1G,2010LNP...817.....B}, is well established although it has a few, but important caveats that have motivated the appearance of alternative theories such as the one discussed in this article. 
Recent up-to-date reviews on the current status of the 
global analysis of data coming from neutrino oscillation and non-oscillation experiments can be found in~\citet{2016NuPhB.908..199G} and~\citet{2018arXiv180409678C}. 
Indeed,
several articles have recently addressed the possibility of the neutrino flavor  oscillations being described by a more general model than the standard model of oscillations in  three  neutrino   flavors.  Two classes of generalizations  have become very popular in the particle physics community: the  possibility of the three neutrinos having non-standard interactions 
with quarks, through a generalized MSW mechanism~\citep[e.g.,][]{2003JHEP...03..011D,2015arXiv150202928P} 
and 3 + 1 (sterile) neutrino flavor  oscillation models~\citep[e.g.,][]{2004PhRvD..69k3002D,2005NuPhB.708..215C}.

In relation to both neutrino models, several constraints have been placed on their fundamental properties using data coming from experimental detectors and astrophysical and cosmological observations. In particular, the solar electron neutrino spectra were computed for both types of neutrino  flavor  models, i.e., for a three-neutrino  flavor oscillation model with a generalized MSW mechanism~\citep{2017PhRvD..95a5023L}, and the  3+1 neutrino   flavor  oscillation model~\citep{2018EPJC...78..327L}. Nevertheless, there is a very important difference between these two class of models: it is the fact that in the 3+1 neutrino   flavor model,  not only will the usual solar electron neutrino spectrum be affected (as well as the  $\mu$- and $\tau$- neutrino spectra), but a new solar sterile neutrino spectrum will be produced in the Sun through  neutrino flavor  oscillations~\citep{2018EPJC...78..327L}.

\smallskip
In this article we are interested in studying  how a particle physics model with a hidden sector comprising dark matter and sterile neutrinos (in which the two sectors are linked) can be probed by the next generation of solar neutrino detectors.
This work uses the 3+1 neutrino model developed by~\citet{2017JCAP...07..021C}. However, in this work, we compute the electron neutrino spectra of all the leading solar neutrino sources for an up-to-date solar model. Specifically, by using an up-to-date stellar evolution code that incorporates the subroutines that accreted dark matter from the halo~\citep{2011PhRvD..83f3521L}, we compute a model in which 
the Sun evolves within a halo of asymmetric dark matter, for which  we then compute the electron neutrino spectra of all the leading solar neutrino sources. 
This neutrino  flavor model corresponds to a generalisation of the 3+1 neutrino flavor  model~\citep{2018EPJC...78..327L}, where the neutrinos can now experience oscillations with the dark matter particles inside the Sun by a new MSW mechanism in the hidden sector. The specific goal of this work is to look for distortions in the shape of  electron neutrino spectra due to the interaction of dark matter particles with sterile neutrinos.   

\smallskip
In this section we have presented a brief review of the arguments in favor of sterile neutrinos being part of the solution to resolve the dark matter problem.  In section~\ref{sec-DMSUN},    we present the basic model to explain the existence of dark matter in the Sun's core.
In section~\ref{sec-Peenu} we present the current model to
study the impact of dark matter in the  3+1 sterile neutrino oscillation model. 
In section~\ref{sec-SNUDMSUM} we  present the predictions 
of solar sterile neutrino spectra  influenced by the presence of dark matter. In the final section~\ref{sec-DC} we discuss the implication of these results for the next generation of solar neutrino experiments.

\begin{figure}
\centering 
\includegraphics[scale=0.45]{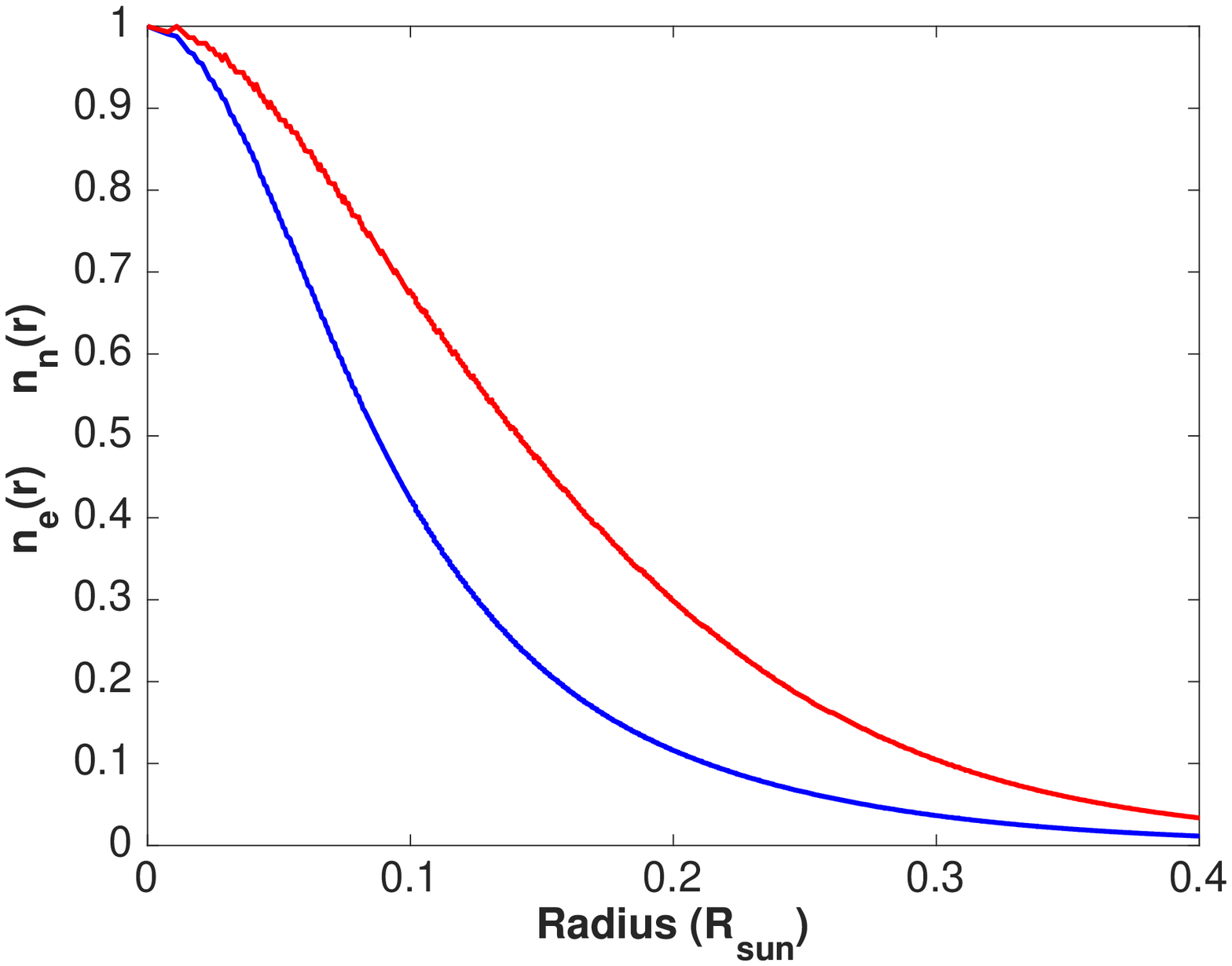}
	
\vspace{0.3cm}
	
\includegraphics[scale=0.45]{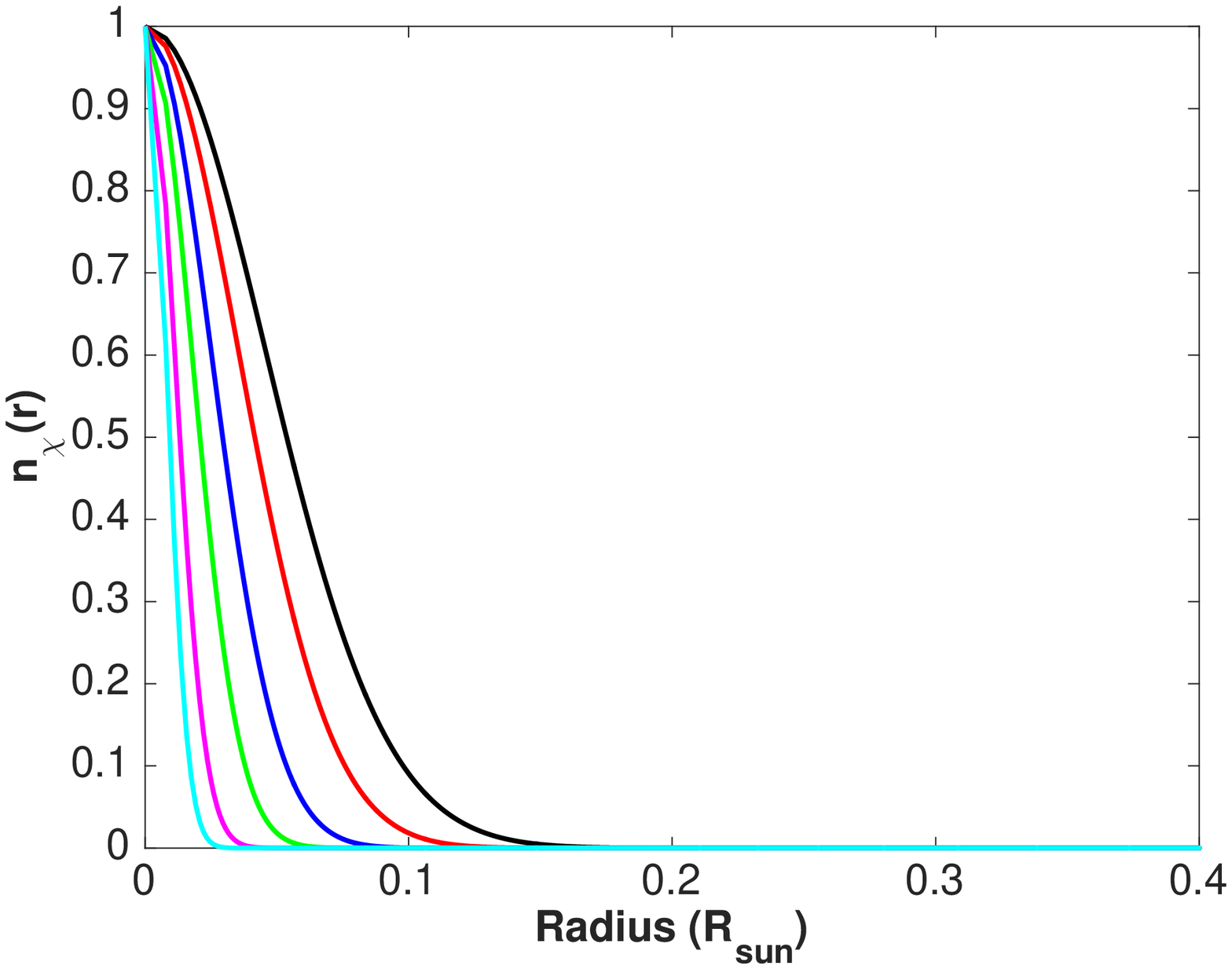}
\caption{Variation of the number densities of electrons,
neutrons, and dark matter particles for the present-day solar model. ({\bf a}) The number densities of electrons $n_e(r)$ (red curve) and neutrons $n_n(r)$ (blue curve). The central 
values of  $n_e$  and  $n_n$ are  $\rm 6.0\times 10^{31}\; m^{-3}$  and 
$\rm 2.7\times  10^{31}\; m^{-3}$. The values can change slightly with the 
properties of the  dark matter halo.   
({\bf b}) The number density of  dark matter particles $n_{\chi}(r)$  for $m_\chi=$ 3 (black curve), 5, 10, 20, 50 and 100 GeV (cyan curve).
To better illustrate the variation of the density functions with radius (in the Sun's core) we fix all central values to 1. For instance, for $n_{\chi}(r)$,  at $\rm r=0$ we have  $n_{\chi}(0)=n_{\chi c}=1$.} \label{fig:1}
\end{figure}


\section{Baryons and dark matter in the Sun}
\label{sec-DMSUN}

The fact that present-day dark matter and baryonic densities are of the same order of magnitude has motivated the idea for
a possible link between the baryonic and dark matter sectors~\citep[e.g.,][]{2014PhR...537...91Z}.
This link can be created by an asymmetry generated in one or both sectors that is then communicated between them, giving rise not only to the baryonic asymmetry, but also to a dark matter asymmetry~\citep[e.g.,][]{2012ApJ...757..130L}. 
In such  models of asymmetry dark matter,  only dark matter particles exist at present day, because  dark matter antiparticles vanished by annihilation with the former in the primitive universe. 
In general, to correctly estimate the number of dark matter particles trapped inside the Sun or a star it is necessary to consider the  self-interaction of dark matter, because a larger number of already accumulated particles may increase the self-capture rate significantly. Dark matter models in the class were studied in the solar case by~\citet{2010PhRvL.105a1301F} and~\citet{2010PhRvD..82h3509T} and extended to other stars by
~\citet[][]{2017PhRvD..95b3507M}. However,  in this study we will be conservative in our analysis, and as such we will only consider the capture of particles due to the interaction of dark matter with baryons. 

\subsection{Capture of dark matter by a star}

As a star travels around the center of our Galaxy, it continuously captures particles from the halo of dark matter that encircles the Milky Way. The star's evolution is more or less affected, depending on the properties of such dark matter particles and the number of them  accumulate in its interior. 
The impact of dark matter on the present structure of the star is regulated by the balance between the capture and annihilation of dark matter particles $\chi$ and antiparticles $\bar{\chi}$ inside it. The process of  dark matter accumulation by the star can be summarized as follows: the dark matter particles of the halo may occasionally scatter with the nuclei of the different chemical elements  present in the star's interior. If such collisions occur between the nuclei and dark matter particles, it may  lead to the dark matter particles  losing enough energy to become trapped in the star's gravitational field,
an to end up by drifting to its center. Such a process that persists overtime ends up by modifying the evolutionary path of a star.   Accordingly, the total number of particles and antiparticles, i.e,  $N_\chi$ and $N_{\bar{\chi}}$, inside the star is determined by the balance between the capture rate and the annihilation rate of dark matter~\citep[e.g.,][]{1996PhR...267..195J}. The values of $N_\chi$ and $N_{\bar{\chi}}$  at a given time of evolution are computed by solving the following system of coupled equations:
\begin{eqnarray}
\frac{dN_i}{dt}=C_i-C_a N_\chi N_{\bar{\chi}},
\label{eq:dNdt}
\end{eqnarray}
where $i$ is equal to $\chi$ (dark matter particles) or $\bar{\chi}$
(dark matter antiparticles). In the above expression 
the evaporation of dark matter particles is neglected. The constant $C_i$ gives the rate of capture of particles (antiparticles) from the dark matter halo
and $C_a$ gives the annihilation rate of particles and antiparticles. In general, the capture and annihilation of dark matter in the star's core are relatively efficient processes; as such the final number of $\chi$ and $\bar{\chi}$ particles accreted by the star depends on the relative magnitude of the  $ C_i $ and $C_a $ processes. Since in this study we are considering only  asymmetric dark matter, this corresponds to assuming that all antiparticles have annihilated with their counterparts in the primitive universe, and only $\chi$ particles have survived until the present day. It follows that the system of coupled equations (\ref{eq:dNdt}) reduces to a single equation with $i=\chi$ and  $C_a=0$, i.e., ${dN_\chi}/{dt}=C_\chi$. Therefore 
over time $N_\chi$ increases in the core of the star, such that  $N_\chi=C_\chi t$.

\subsection{Evolution of the Sun in a halo of dark matter}

The total number of dark matter particles present in the Sun's core 
is calculated using a well established code to study the evolution 
of a star in a halo of dark matter.  
In this code the mechanism for the capture
 and energy transport of dark matter
 were implemented as explained in~\citet{2011PhRvD..83f3521L}. This code was built by
originally modifying  the stellar   evolution code
 {\sc Cesam}~\citep{2008Ap&SS.316...61M}, a sophisticated 
 code used to compute high-quality solar models.
 In particular, the code uses the equation of state and opacity coefficients from~\citet{1996ApJ...464..943I}, a Hopf solar atmosphere,
and an adapted low-temperature opacity table. These models also included the appropriate screening, for which  the Mitler prescription~\citep{1995ApJ...447..428D} is used, and also take into account
the microscopic diffusion of helium and other chemical elements
~\citep{1998ApJ...506..913B}. See~\citet{2004PhRvL..93u1102T,2010ApJ...715.1539T} and references therein for further details.  The calculation of the solar models (with or without dark matter) is done using a set of input parameters that follow the well established criteria  used to compute a standard solar model~\citep{1993ApJ...408..347T}.  Each solar model evolves from the pre-main sequence until the  present age of 4.5 Gyr. Typically in a scenario where a significant amount of dark matter is captured by the star,  a single evolution model can have more than 140 time steps until it  reaches the present age. Each solar model at a given epoch  has more than 2000 layers.  The stellar code requires that  the structure equations are solved with an accuracy of $10^{-5}$.  
In particular, the results of our modified solar model \citep[e.g.,][]{2011PhRvD..83f3521L} are in agreement with those of other codes in the literature~\citep[e.g.,][]{2016JCAP...11..007V}. 

\smallskip
All solar models with or without dark matter are computed 
for an updated solar mixture of~\citet{2009ARA&A..47..481A}
by  adjusting the initial helium content and the mixing length parameter in such a way that at the present age of  Sun~\citep{2013MNRAS.435.2109L} all the models reproduced the observed solar mass, radius, luminosity and $Z/X$ surface abundance. Equally the solar model without dark matter is required to have an acoustic seismic spectrum and solar neutrino fluxes similar to other models found in the literature~\citep{2010ApJ...715.1539T,2011ApJ...743...24S}.
Moreover,  the reference solar model used in our work is in full agreement with the standard picture of stellar evolution~\citep[e.g.,][]{1993ApJ...408..347T,2010ApJ...715.1539T,2011ApJ...743...24S}. 

\smallskip
We have computed the evolution of the Sun in several dark matter particle scenarios, where $m_\chi$ takes the values 3, 5, 10, and 20 GeV. In agreement  with recent constraints for  spin-independent scattering cross section of the interaction of dark matter with baryons $\sigma_{\rm SI}$ , we choose the values $10^{-37}\; {\rm cm^2}$ for $m_\chi \sim 3\, {\rm GeV} $ and $10^{-41}\; {\rm cm^2}$  for $10\, {\rm GeV} \le m_\chi \le 100 \, {\rm GeV}$~\citep{2017PhRvL.118y1301A}. Similarly  for the constraints on the spin-dependent scattering cross section of dark matter with baryons $\sigma_{\rm SD}$, we choose  the values  $10^{-34}\; {\rm cm^2}$
for $m_\chi \sim 3\,{\rm GeV}$ and $10^{-44}\; {\rm cm^2}$  for
$10 \le m_\chi \le 100 \, {\rm GeV} $~\citep{2017PhRvL.118y1301A}.
The other parameters used to define the properties of the dark matter halo can be found in~\citet{2011PhRvD..83f3521L}; in particular, for the  density of the dark matter we choose  $\rho_{\chi}\sim 0.38\; {\rm GeV\, cm^{-3}}$. These numerical simulations have produced several solar models for the present day, for which $N_\chi$ has taken values between $10^{37} $ and $10^{47}$. The largest values correspond to a solar model with a large $\sigma_{\rm SI}$ value. 

\begin{figure}[!t]
	\centering 
	\includegraphics[scale=0.40]{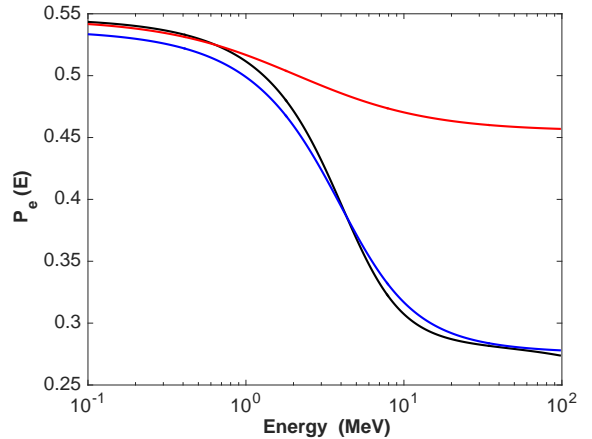}
	\caption{The survival probability of electron neutrinos: the $P_{e}$ curves correspond to three neutrino models: three-flavor neutrino oscillation model (classical model; black curve); 3+1 flavor neutrino oscillation model (without dark matter; blue curve); 3+1  flavor  neutrino oscillation model and dark matter 
		with an effective strength of  $G_\chi=2.8\times 10^{10}\; G_{\rm F}$
		(red curve). 
	}
	\label{fig:Pe}
\end{figure}  

\begin{figure}[!t]
\centering 
\includegraphics[scale=0.45]{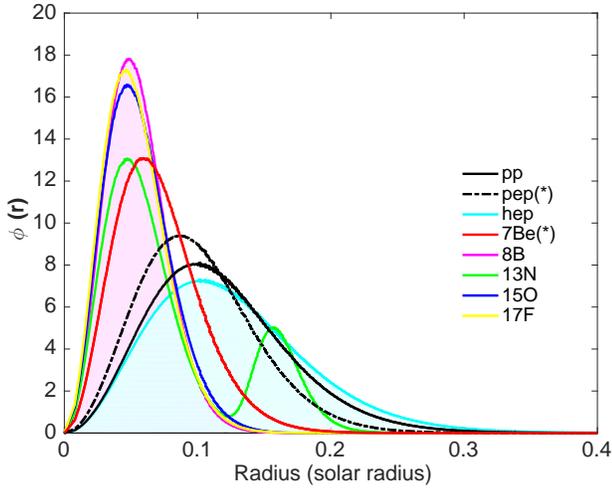}
\caption{
Neutrino emission sources of the various nuclear reactions of the proton-proton (pp) chain and carbon-nitrogen-oxygen (CNO) cycle present inside the Sun. The standard solar model used in this calculation is in good agreement with the helioseismology data, and its internal structure is identical to many other solar models found in the literature (see text). The figure shows
the variation of the neutrino source $\phi_i(r)$ ($i=$, $\rm pp$, $\rm  pep(*)$, $\rm  hep$, $\rm ^8B$, $\rm ^7Be(*)$, $\rm ^{13}N$, $\rm ^{15}O$ and $\rm ^{17}F$) with the solar radius $r$. All the solar neutrino sources produce a continuous spectrum,
with the exception of the $\rm pep(*)$ and $\rm ^7Be(*)$ sources that generate spectral lines.  The red and blue shaded areas correspond to the  $\rm ^8B$
and $\rm hep$ sources,  which are the most sensitive to the hidden sector (see text).
Although solar models with dark matter will show a slight variation of the $\phi_i(r)$ functions, overall the $\phi_i(r)$ for solar models
 with and without dark matter will have similar. This figure was adapted from~\citet{2017PhRvD..95a5023L}.
}
\label{fig:Phinu}
\end{figure}

\begin{figure}[!t]
\centering 
\includegraphics[scale=0.45]{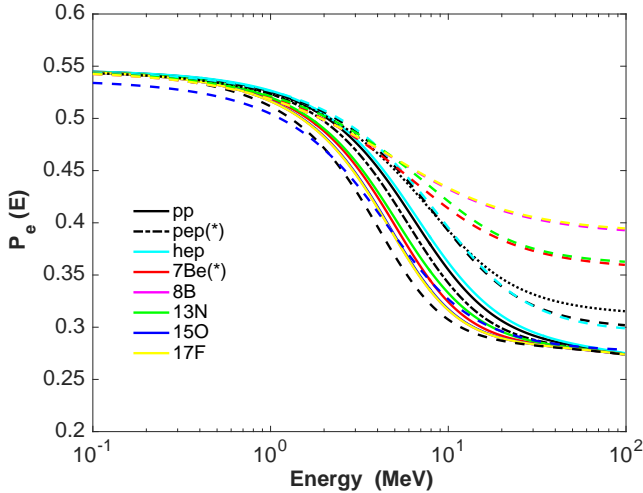}
\caption{The variation of the different electron neutrino survival probability functions ($P_e$) with the energy of the emitted neutrino
(the color scheme is the same one as in Figure~\ref{fig:Phinu}): the continuous and dashed curves correspond to the standard model of three active neutrinos and to the 3+1 neutrino model with dark matter (see Figure~\ref{fig:Pe}). The $P_e$ functions of the $\rm pep(*)$ neutrino source are shown as  black curve: a dotted-dashed line for the standard neutrino model and  a dotted line for the  3+1 neutrino models.}
\label{fig:PePhinu}
\end{figure}

\subsection{Electrons and neutrons} 

The propagation of neutrinos in matter is strongly dependent 
on the local distributions of electrons and neutrons. Figure~\ref{fig:1} shows the density of electrons $n_e(r)$ and the density of neutrons $n_n(r)$  inside the Sun at the present age as predicted by an up-to-date standard solar
model~\citep{2013MNRAS.435.2109L}. 
The fact that  $n_n(r)$ is much larger in the Sun's center than at its surface results from the fact that more than 70\%  the Sun's core at the present day consists $^4He$.  Consequently, the  Sun's core has a larger number of neutrons  than the more external layers, leading to a decrease in $n_n(r)$ from the center towards its surface. This important point is illustrated in Figure~\ref{fig:1}, which shows the relative density of electrons and neutrons inside the Sun.  These quantities have a major impact on the propagation of neutrinos, as we will discuss in the later sections. It is worth noticing that these quantities vary only slightly between solar models computed for different types of dark matter particles; the general form of these curves remains the same. 

\begin{figure}[!t]
	\centering 
	\includegraphics[scale=0.45]{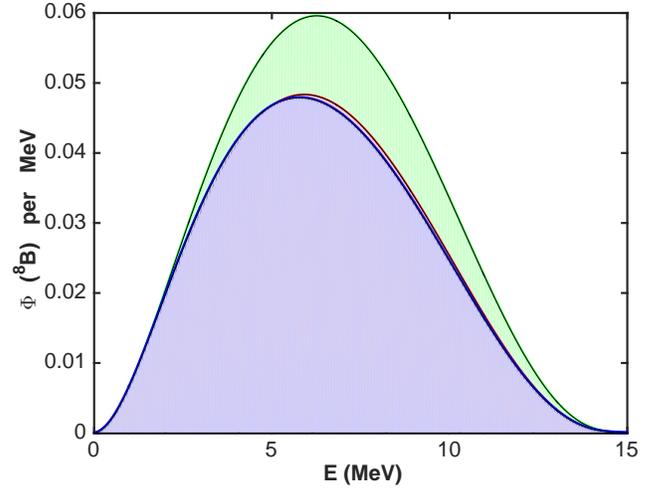}
	\caption{
 	$\Phi_{{^8B},\odot} (E)$ is the electron neutrino spectrum for the present Sun in one of the following flavor neutrino oscillation models:
		({\bf a}) standard (three active neutrino) neutrino model (blue area);
		({\bf b}) 3+1 sterile neutrino model (red area);
		({\bf c}) 3+1 sterile neutrino model with dark matter 
		with an effective strength of  $G_\chi=2.8\times 10^{10}\; G_{\rm F}$
		(green area). In the calculation of these neutrino spectra 
		we used an up-to-date standard solar model.
	}
	\label{fig:PeB8}
\end{figure}  


\begin{figure}[!t]
\centering 
\includegraphics[scale=0.45]{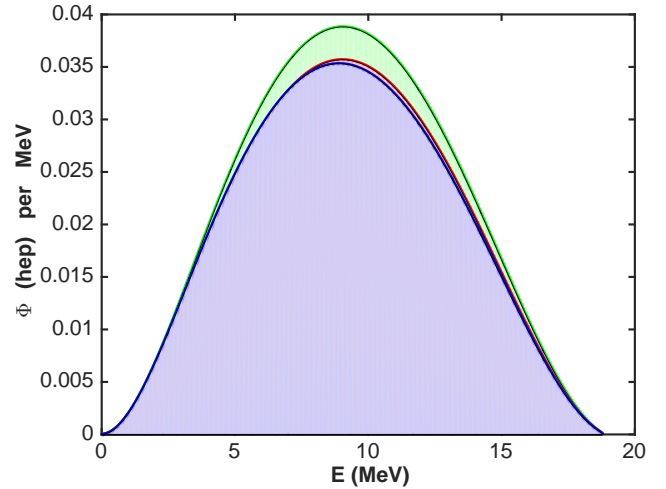}
\caption{The electron neutrino spectrum of the $\rm hep$		
neutrino source ($\Phi_{{hep},\odot} $) for the present Sun. 
Th color scheme is the same  as in Figure~\ref{fig:PeB8}.}
	\label{fig:Pehep}
\end{figure}

\subsection{Dark matter particles}

The accumulation of dark matter particles in the Sun's core during its evolution from the pre-main-sequence phase until its present age
not only affects  the structure of the core but, more importantly
for this study, it also determines the  number  of dark matter particles $N_\chi$ today. Nevertheless, the distribution of dark matter in the Sun's core is
well approximated  by the expression $ n_\chi(r)=n_{\chi c}\exp{(-\frac{r^2}{r^2_\chi})} $ where $ n_{\chi c}$ is the central density of dark matter particles and  $r_\chi $ is the characteristic radius of the dark matter core. The quantity  $ n_{\chi c}$ depends on the total 
number of particles captured by the Sun at a fixed 
time $N_\chi$ (as given by Equation~\ref{eq:dNdt}), for which $n_{\chi c}={N_\chi}/{(r_\chi^3\pi^{3/2})} $ with $r_\chi=\sqrt{{3T_{c\odot}}/{(2\pi G_N \rho_{c\odot} m_\chi)}} $ where $T_{c\odot}$ and $\rho_{c\odot}$ are the central temperature and density of the Sun.  Two quantities affect  $n_{\chi}(r)$ significantly:  the values of $N_\chi$ and  $r_\chi$, both of which depend on the evolution of the Sun and can be obtained from the numerical simulations.  
Figure~\ref{fig:1}  shows the distribution of dark 
matter particles in the Sun's core for dark matter particles with different masses.  It is quite clear that the more massive particles are concentrated in the Sun's core.

\section{Propagation of sterile neutrinos inside the Sun}
\label{sec-Peenu}

In the generalization of the standard model of neutrino  flavour  oscillations, i.e., that with three active neutrino flavors, any number of sterile neutrinos are allowed to exist~\citep[e.g.,][]{2011PhRvD..83k3013P}. Nevertheless, even the simplest of these generalizations, in which the standard neutrino   flavour model is extended to include only one sterile neutrino $\nu_s$, i.e. the 3+1  sterile neutrino model $(\nu_e,\nu_{\tau},\nu_{\mu},\nu_s)$,  is sufficient to explain the neutrino anomalies mentioned previously~\citep{2013JHEP...05..050K}. In this case only two additional parameters are needed to  define the flavor oscillation of $\nu_s$: 
$|\Delta m^2_{41}|$ (the mass spilling)  and $\theta_{14}$ (the mixing angle).

\smallskip
\citet{2013PhRvD..88g3008G} have found that a global analysis of all the neutrino oscillation data sets can be fitted by a unique 3+1 neutrino model. These authors have found that in this case $\Delta m^2_{41}$ takes a value between  $0.82$ and $ 2.19\; {\rm eV^2}$.  This result was confirmed by an independent analysis of ~\citet{2017PhRvL.118l1802K} in which $|\Delta m^2_{41}|$ has similar values, i.e., $|\Delta m^2_{41}|$ varies between $0.2$ and $2.3\;{\rm eV^2}$.  
Similarly~\citet{2015PhRvD..91i5023B} and~\citet{2017PhRvL.118l1802K} found for the upper limit of $\theta_{14}$ the values: $10.6^o(0.1855)$ and  $9.2^o(0.1582)$.  

\smallskip
We could expect that the presence of dark matter in the Sun's core might affect the global analysis of neutrino oscillation data. Nevertheless, this effect is actually relatively small, since this global analysis depends marginally on the properties of the matter (by the MSW effect) in the core of the Sun. Therefore, even if the impact 
dark matter is such that it slightly affects the parametrization of the neutrino  flavor model adopted, the main conclusions of this work will not change.  

\smallskip
Accordingly, for this study we choose to adopt as fiducial parameters
 the values obtained by different authors  by fitting the observational neutrino data to a 3+1 flavor  neutrino oscillation model: $\Delta m^2_{21}= (7.5\pm 0.14)\times 10^{-5}{\rm eV^2}$,
$\sin^2({\theta_{12}})=0.300\pm 0.016 $, and $\sin^2{\theta_{13}}=0.03$
from~\citet{2015PhRvD..91i5023B};
$\Delta m^2_{31}= (2.457\pm 0.047)\times 10^{-3}{\rm eV^2}$ 
from~\citet{2016NuPhB.908..199G}; and $\Delta m^2_{41}=1.6 \ {\rm eV^2}$, $\theta_{14}=0.1186$,  $\theta_{24}=0.1651$,  $\theta_{34}=0.000$,  and all $\delta_{13,14,34}=0$ from~\citet{2017JCAP...07..021C}. 
This last set of parameters is  favored by  short-baseline neutrino data. 
Moreover, we tested the sensitivity of the neutrino model to the parameters
$\Delta m^2_{21}$ and ${\theta_{12}}$ by adopting the values found by~\citet{2017JCAP...07..021C}, and as expected the results found were 
identical to the previous ones.

\subsection{The survival probability of electron neutrinos}

The objective of this work is to compute the imprint 
of dark matter via sterile neutrinos on the spectrum of 
electron neutrinos produced by the Sun, specifically to 
identify how the shape of the neutrino spectrum changes due to the presence of dark matter in the Sun's interior.  Fortunately, for the range of parameters we are interested in, the propagation of neutrinos from the Sun's core to the Earth  takes a particularly simple form, since the evolution of neutrinos in space is identical
to neutrino propagation in vacuum, and the propagation in matter is adiabatic. For that reason it is possible to compute the number of electron neutrinos arriving on Earth  using a relatively simple formula~\citep[e.g.,][]{2010LNP...817.....B}. 

The evolution of the 3+1 neutrino flavor eigenstates
$(\nu_e,\nu_{\mu},\nu_\tau,\nu_s)$ is governed by a Schr\"odinger-like equation. Following~\citet{1989RvMP...61..937K}
conveniently, away from resonances, it is possible to compute the flavor oscillation by assuming a simplified picture of flavor
oscillation of one sterile neutrino, $\nu_s$, and an active one, which we chose to be $\nu_e$. The motivation for such decomposition into two neutrino flavors   is also discussed in~\citet{2018EPJC...78..327L}.
This two-flavor neutrino model  $(\nu_e,\nu_s)$ is sufficient to explain the impact of the dark matter in the 3+1 neutrino flavor oscillation model~\citep[e.g.,][]{2011PhRvD..83k3013P}. Among the several 
two-flavor  neutrino models available in the literature  that have been proposed to approximately compute the electron probability  in the 3+1 neutrino model~\citep[e.g.,][]{2000NuPhB.583..260L,2015PhLB..744...55M}, 
we  choose the model obtained  by~\citet{2017JCAP...07..021C}, 
which is in good agreement with the numerical solution.
Thus, for a neutrino energy E, the evolution equation can be written as
\begin{eqnarray}
i\frac{d}{dx}
\left[ {\begin{array}{c}
\nu_e  \\
\nu_s \\
\end{array} } \right]=
\left[ {\begin{array}{cc}
V_x-\Delta {c}_{2\theta} \qquad & 	V_y+\Delta {s}_{2\theta}\\
V_y^*+\Delta {s}_{2\theta} \qquad &		-V_x+\Delta {c}_{2\theta} \\
	\end{array} } \right]
\left[ {\begin{array}{c}
	\nu_e  \\
	\nu_s \\
	\end{array} } \right],
\end{eqnarray}
where  $\Delta {c}_{2\theta}=(\Delta m_{21}^2/4E)\cos{(2\theta_{12})}$ and $\Delta  {s}_{2\theta}=(\Delta m_{21}^2/4E)\sin{(2\theta_{12})}$,
and $V_x$ and  $V_y$ are functions related to the 
effective matter (ordinary and dark matter) potential inside the Sun.
If we keep only the first-order terms, the survival probability $P_{e}$  is
\begin{eqnarray}
P_{e} =
s^4_{13}+c^4_{13}c^4_{24}s^4_{14}+a_m+b_m, 
\label{eq:Pee}
\end{eqnarray}
where $c_{ij}=\cos{\theta_{ij}}$, $s_{ij}=\sin{\theta_{ij}}$,
and $a_m$ and $b_m$ are dependent on the internal
structure of the Sun. The functions $a_m$ and $b_m$ 
are given by the following expressions:
\begin{eqnarray}
a_m=c^4_{13}\left(c_{14}s_{12}-c_{12}s_{14}s_{24}\right)^2
\left(c_{14}s_{m}-c_{m}s_{14}s_{24}\right)^2
\nonumber
\end{eqnarray}
and
\begin{eqnarray}
b_m=c^4_{13}\left(c_{14}c_{12}+s_{12}s_{14}s_{24}\right)^2
\left(c_{14}c_{m}+s_{m}s_{14}s_{24}\right)^2,
\nonumber
\end{eqnarray}
with $c_{m}=\cos{\theta_{m}}$ and $s_{m}=\sin{\theta_{m}}$,
where $\theta_{m}$ is the matter angle, which depends on the  local distribution of baryons and dark matter particles inside the Sun.  The function $\theta_{m}$ is given by
\begin{eqnarray}
\cos{(2\theta_{m})}=\frac{{\cal M }_x}{
	\sqrt{{\cal M }_y^2+{\cal M }_x^2}},
\label{eq:thetam}
\end{eqnarray}
with ${\cal M }_x\equiv \cos{(2\theta_{12})} -V_x$ and ${\cal M}_y\equiv |\sin{(2\theta_{12})} +V_y|$ where $V_x$ and 
$V_y$ are functions related to the effective matter potential.

\subsection{The effective matter potential}

Neutrinos propagating throughout the Sun's interior will change
flavor due to vacuum oscillations and matter oscillations.  
As the solar plasma is now composed of  neutrons, protons, electrons,  and dark matter particles, some of these particles will contribute to the neutrino flavor  oscillation mechanisms.  The interaction of neutrinos with the solar medium proceeds through coherent forward elastic charged-current (${\rm cc}$) and neutral-current (${\rm nc}$) scatterings; as usual, these effects are represented by the effective potentials  $V_{{\rm cc}}$ and $V_{{\rm nc}}$. These potentials  are expressed as fu{\rm nc}tions of local number densities $n_e(r)$ and $n_n(r)$:  
\begin{eqnarray}
V_{{\rm cc}} =\sqrt{2}G_{\rm F}n_e(r)\eta_{\nu}
\end{eqnarray}
and 
\begin{eqnarray}
V_{{\rm nc}} =\sqrt{2}G_{\rm F}\frac{1}{2}n_n(r)\eta_{\nu},
\end{eqnarray}
where $G_{\rm F}$ is the Fermi constant and  $\eta_{\nu}$ is the ratio of the neutrino energy $E$ to  $\Delta m^2_{21}$ given by $\eta_{\nu} (E)=4E/\Delta m_{21}^2$.  $\Delta m^2_{21}$ 
is computed as $\Delta m^2_{i1}=m^2_{i}-m^2_{1}$ ($i=2,3,4$).
Similarly, $V_{\chi}$ is the potential related to dark matter
particles~\citep{2017JCAP...07..038F}:
\begin{eqnarray}
V_{\chi}(r) = \;G_\chi \; n_{\chi }(r)\;\eta_{\nu},
\end{eqnarray}
where $ G_\chi=\alpha_\chi{g_{\rm B}^2}/{m_{\rm B}^2}$. 
This latter quantity is the equivalent of the Fermi constant in the 
hidden sector, for which $g_{\rm B}$ and $m_{\rm B}$ are the coupling constant
and the mass of the hidden boson $B$, $\alpha_\chi$ is a factor of
the order of unity, which we will set to be one.  In the calculation of $V_\chi$ we choose $G_\chi=2.8\times 10^{10} G_{\rm F}$. This value comes from a constraint obtained from the CMB data~\citep{2017JCAP...07..038F}.

The potential $V_\chi$ makes  a significant impact on
the 3+1 neutrino  flavor   model. This term introduces
the mechanism by which dark matter can affect the
variation of neutrino flavor. Its strength is regulated by the
magnitude of  $ G_\chi$.

The contributions of the effective matter potentials
$V_{{\rm cc}}$ and $V_{{\rm nc}}$ and dark matter potential $V_{\chi}$  
are taken 
into account in the previous Equation~\ref{eq:thetam}
thorough the  expressions $V_{x}$ and $V_{y}$: 
\begin{eqnarray}
V_x=\frac{1}{2}\left[V_{{\rm cc}} c^2_{13}
\left(c^2_{14}-  s^2_{14}s^2_{24}\right)
+ V_s \left(s^2_{14}-  c^2_{14}s^2_{24}\right) \right]
\end{eqnarray}
and
\begin{eqnarray}
V_y=\left(V_s -V_{{\rm cc}} c^2_{13} \right)c_{14} s_{14} s_{24}, 
\end{eqnarray}
where $V_{s}=V_{\chi} -V_{{\rm nc}}$. 

\smallskip
The calculation of $V_\chi$ will be done for a specific model of dark matter. Since there is a large variation in the total number of dark matter particles with the properties of the dark matter, we choose the fiducial extreme value of    $N_{\chi}\sim 10^{47}$ to illustrate the impact of dark matter on the 3+1 neutrino  flavor   model.  

\subsection{The survival of electron neutrinos}

As in the classical neutrino flavor  oscillation model, the conversion of electron neutrinos due to neutrino  flavor  oscillations depends on the local electron density $n_e(r)$~\citep{2013ApJ...765...14L}, but this generalized neutrino flavor  oscillation model also depends on the local neutron density $n_n(r)$ and  the local dark matter density $n_\chi(r)$. 

Figure~\ref{fig:Pe} shows the variation of the $P_e(E)$ with the energy of the neutrino. $P_e$  is given by Equation~\ref{eq:Pee}; the values of $P_e$ are particularly accurate in the case that $s_{34} = 0$ and when ${V_{{\rm cc}}E}/{\Delta m^2_{31}}\ll 1$
\citep{2017JCAP...07..021C}.
 It is worth noticing that in relation to the standard model of three active neutrino model, the 3+1 model (without dark matter) affects equally the vacuum and matter of the electron neutrino, and the  3+1 model (with dark matter) affects most of the matter oscillations. In the latter model this effect is quite significant. The reason is related to the large value of $G_\chi$ adopted in these calculations. 
Indeed, while the standard model three active neutrinos 
and the 3 +1 (without dark matter) neutrino model~\citep{2018EPJC...78..327L}
 have very similar survival electron probability $P_e$ curves
(compare black and blue curves), the 3 +1 (with dark matter) neutrino  flavor  model has a  $P_e$ (red curve) clearly distinct from the previous models. This difference is more pronounced  for neutrinos of higher energy (see Figure~\ref{fig:Pe}).
An important aspect of this result worth highlighting is 
 that the $P_e(E)$ sterile component (caused by the  new parameters) affects electron neutrinos with any energy equally. However, the $P_e(E)$ dark matter component (related to the dark matter potential $V_\chi$) affects mostly the neutrinos with higher energy.  This last term is very sensitive to $G_{\rm F}$ and the $N_\chi$.

\section{The solar electron neutrino spectra}
\label{sec-SNUDMSUM}

The solar electron neutrinos arriving on Earth are produced  by the nuclear reactions of  the proton--proton (PP) chains and the carbon-nitrogen-oxygen (CNO) cycle at very high temperatures in the deepest layers in the Sun's interior. Therefore to compute the neutrino spectra, we must take into account the location of neutrino sources inside the Sun. Thus to calculate the average survival probability of electron neutrinos for each nuclear reaction in the solar interior, i.e., $P_{e,i} \equiv \langle P_{e} (E)\rangle_i $, we compute
\begin{eqnarray} 
P_{e,i} = 
A_i^{-1} \int_0^{R_\odot} P_{e} (E,r)\phi_j (r) 4\pi \rho(r) r^2 dr, 
\label{eq:Pnuej}
\end{eqnarray}  
where  $ A_i $ is a normalization constant given by  $A_i= \int_0^{R_\odot}\phi_i (r) 4 \pi \rho (r) r^ 2  \;dr $, 
and $ \phi_i (r) $  is the electron neutrino emission function for the  nuclear reaction $i$. $i$ corresponds 
to the following electron neutrino nuclear reactions: ${\rm pp}$, ${\rm pep}$, ${\rm ^8B} $, ${\rm  ^7Be}$, ${\rm ^{13}N} $, ${\rm  ^{15}O}$ and ${\rm ^{17}F} $  (see figure~\ref{fig:Phinu}).
A detailed discussion about the neutrino sources inside
the Sun, and the properties of these specific solar models can be found in~\citet{2013PhRvD..88d5006L}. 

\smallskip
Figure~\ref{fig:Phinu}  shows the local source of electron
neutrinos for the standard solar model, which by definition is consistent with the current helioseismic and solar flux data. 
A detailed account of the properties of Equation~\ref{eq:Pnuej}
can be found in~\citet{2017PhRvD..95a5023L}. It is worth noticing that in general the dark matter affects significantly the evolution of the star in its core. Although  the location of the different  neutrino sources 
can vary with the  dark matter, its effect on $\phi_i(r)$ is relatively small
(see Figure~\ref{fig:PePhinu}). 
The electron neutrinos coming from the different nuclear reactions have a very similar behavior in terms of energy: neutrinos with relatively low energy are only affected by vacuum oscillations, while neutrinos with high energy are affected by vacuum and matter oscillations. 
Both  of these effects are taken into account by the $P_{e}(E)$ function (Equation~\ref{eq:Pee}). Moreover, since neutrinos are produced in nuclear reactions located at different solar radii, the $P_{e,i}(E)$ curves vary slightly due to $\phi_i$ (Equation~\ref{eq:Pnuej}).  This is the reason why in the standard neutrino  flavor  model the $ P_{e,i} (E) $ are very similar for low- and high-energy neutrinos, with only small differences. However, if dark matter is present, while the difference is for the low-energy neutrinos,  the effect is very large for very high-energy neutrinos. This is due to the contribution of $V_\chi(r)$  to $P_{e}(E)$ (see Figure~\ref{fig:PePhinu}).  Actually, this is the reason why one could expect that neutrinos produced in nuclear reactions located near the center of the Sun, such as the neutrino sources $\rm ^8B$ and $\rm ^7Be$ (PP chains) and $\rm ^{13}N$, $\rm ^{15}O$ and $\rm ^{17}F$ (CNO cycle), to be very affected by the  presence of dark matter particles.
Nevertheless, as we will discuss  next, this is not necessary the case, because the neutrinos more affected are the ones produced near the Sun's center, but also the ones that have a relatively high energy.

\smallskip

Now that we have all the ingredients, then we can compute the electron neutrino energy spectrum of the different solar sources. The energy spectrum of electron neutrinos produced by a specific nuclear reaction in the Sun is essentially independent of the properties of the solar plasma at the location where occur nuclear reaction occurs. With a great degree of certainty we will consider that this neutrino spectrum  is similar to the neutrino spectrum measured on  laboratory on Earth for the same nuclear reaction. A typical example is the $\rm ^8B$ spectrum~\citep[e.g.,][]{2006PhRvC..73b5503W}.
Accordingly, we can use such an Earth spectrum  as a reference against which we can test the new physics of our model. 
Hence, the difference observed between the neutrino spectrum given by the nuclear reaction in the Sun's core, $\Phi_{i}(E)$,  and the neutrino spectrum that emerges from the Sun's surface and is measured by a solar neutrino detector on Earth, $\Phi_{i,\odot}(E)$ is only due to neutrino  flavor  oscillations
(either vacuum oscillations or vacuum-matter oscillations). Hence, the two previous neutrino energy spectra are simply related:
\begin{eqnarray} 
\Phi_{i,\odot}(E)=P_{e,i} (E) \Phi_{i}(E).
\label{eq-Psie}
\end{eqnarray} 

Figures~\ref{fig:PeB8} and~\ref{fig:Pehep}  show two of these spectra. We notice that although several neutrino sources are located near the Sun's center, only the spectra of $\rm ^8B$ electron neutrinos and $\rm  hep$ electron neutrinos have their spectra modified by the presence of dark matter. This occurs because both sources produce neutrinos with relatively high energy. The effect is more pronounced in the $\rm ^8B$ electron neutrino
spectrum because this neutrino source is located very near the core.
The spectra of the other neutrino sources -- $\rm  pp$, $\rm  pep$ and $\rm  ^7Be$ (pp chain reactions) and $\rm ^{13}N$, $\rm  ^{15}O$  and $\rm  ^{17}F$ (CNO cycle reactions) -- are not shown because the effect of dark matter is negligible, i.e., these spectra are not affected by the presence of dark matter in the Sun's core. 

\begin{figure}[!t]
\centering 
\includegraphics[scale=0.45]{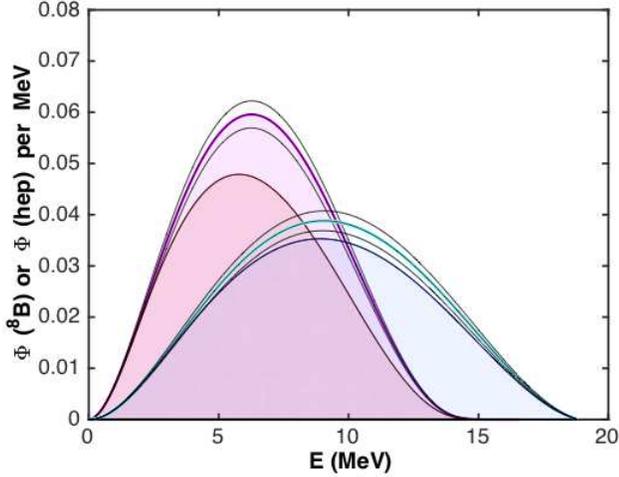}
\caption{
$\Phi_{{^8B},\odot} (E)$  (light red area)  and $\Phi_{{\rm hep},\odot} (E)$  (light blue area) are the electron solar neutrino spectra for the
3+1 neutrino model with dark matter. The error bars 
are shown as  thin black lines around the $\Phi_{{^8B},\odot} (E)$   and $\Phi_{{\rm hep},\odot}$ curves. These black lines were computed assuming that $P_e(E)$ used to compute each of the $\Phi_{i,\odot} (E)$ spectra  has an error of $0.1$. This corresponds to the error 
associated with the  electron survival probability of $P_e(E)=0.29\pm 0.1$ for an $\rm ^8B$ neutrino of energy $8.9$ MeV~\citep{2010PhRvD..82c3006B}.
For comparison, the $\Phi_{{^8B},\odot} (E)$  (dark red area)  and $\Phi_{{hep},\odot} (E)$ (dark blue area)   curves that correspond to the three-neutrino model flavor model are also shown. 
Notice that the $\Phi_{{^8B},\odot} (E)$   and $\Phi_{{hep},\odot} (E)$  for the three neutrino model flavor  model 
are very similar to the ones corresponding to the 3+1 neutrino flavor  model (without dark matter), 
as shown in Figures~\ref{fig:PeB8} and~\ref{fig:Pehep}.
For clarity we have not included the error
bar in the standard neutrino model.
}
\label{fig:Pe_error}
\end{figure}  

Figure~\ref{fig:Pe_error} shows the error bars
of the $\rm ^8B$ and $\rm hep$ spectra computed by assuming a realistic experimental error on the determination of these neutrino spectra.
Accordingly, the error bars used in this calculation correspond to the 
maximum error estimated from the experimental errors on the neutrino fluxes computed by the different  solar experiments, Borexino, SuperKamiokande, and SNO. Assuming the $\rm ^8B$ neutrino flux predicted by the high-metallicity standard solar model as used in this work, this corresponds to an  electron survival probability  of  $P_e(E)=0.29\pm 0.1$ for a neutrino of energy $8.9$ MeV~\citep{2010PhRvD..82c3006B}. The final error bars also include uncertainty coming from the mixing angles  and mass differences, as well as the theoretical uncertainty coming from the standard solar model. Moroever, the function $P_e(E)$ is only weakly sensitive to the physics of the solar model and is mostly affected through the variation of the radial distribution of electrons $n_e(r)$ or the radial density profile $\rho(r)$, which due to helioseismology
is a quite robust quantity in the standard solar model. Therefore, the current uncertainties in the physics of the solar interior, such as the solar abundance metallicity problem, have a small impact in $P_e(E)$.  

It follows from our analysis that in the case of the $\rm ^8B$
spectrum it is such that it is possible to distinguish between 
the 3+1 neutrino model (with dark matter) and the standard model of
three active neutrinos. Therefore it is possible to distinguish between the two
models. In the case of the $\rm  hep$ neutrino spectra the distinction between
the two models is not possible with the current set of data. Finally, we highlight the forthcoming next generation of neutrino experiments such as  LENA~\citep{2012APh....35..685W},
for which a data set corresponding to five years of measurements 
will allow us  to obtain an error four times smaller, i.e., $P_e(E)=0.29\pm 0.025$, for which these constraints will be much stronger.


\section{Discussion and conclusion} 
\label{sec-DC}

In this study, we have computed the expected alteration of the shape of the solar neutrino spectra expected to occur in a 3+1 flavor  neutrino oscillation model due to the existence of dark matter in the Sun's core.
This new type of interaction, mostly due to the oscillations induced by
 dark matter as a result of a matter (MSW) oscillation mechanism in the hidden sector, depends on the specific properties of the dark matter particles, and also on the parameters of the neutrino flavor  oscillation model. The most important factors affecting the neutrino flavors are the concentration of dark matter in the Sun's core and the Fermi coupling constant of the hidden sector. There is also a small dependence on  the local thermodynamic properties of the Sun's core.
\smallskip

Nevertheless, it is worth mentioning that the solar neutrino spectrum for the 3+1 neutrino flavor oscillation model with dark matter presents significant differences from the standard three-neutrino flavor  oscillation model, as well as from a non-standard  neutrino flavor  oscillation model such as the three-neutrino flavor  oscillation model with a generalized MSW mechanism~\citep{2017PhRvD..95a5023L}
and  of the  3+1  neutrino flavor  oscillation model~\citep{2018EPJC...78..327L}.
The neutrino  flavor  model discussed in this work is distinguished from
the previous models mainly by the combination of two factors:
the impact on the neutrino fluxes is limited to the Sun's core, where  the accumulation of dark matter occurs, and only neutrinos with a large energy 
are affected by neutrino  flavor   oscillations. It is the combination of these two effects that causes the 3+1 neutrino  flavor   model with dark matter to have a pronounced impact on the  $\rm hep$ and $\rm ^8B$ neutrino spectra. This behavior is markedly different from the  spectral variations found in the neutrino flavor  oscillation model with a generalized MSW mechanism~\citep{2017PhRvD..95a5023L} and the 3+1 neutrino flavor  oscillation model~\citep{2018EPJC...78..327L}.  In the first case, all the  neutrino sources are affected. However, in the second case,   an impact on the solar electron neutrino spectrum is predicted, as well a new solar sterile neutrino spectrum.
Since dark matter particles inside the Sun are located very near the center, and only electron neutrinos emitted by the nuclear reactions with highest energy are affected, then the neutrino spectrum most modified is the spectrum emitted by the $\rm ^8B$ nuclear reaction. The neutrino spectrum emitted by the $\rm hep$ nuclear reaction is
similarly affected, although the effect is a smaller than in previous example. The impact of dark matter on the spectra of the other nuclear reactions is negligible.

\smallskip
Recently,~\citet{2017JCAP...07..038F} used CMB observational data were able to constrain the Fermi coupling constant $G_{\chi}$ for the interactions of dark matter particles with sterile neutrinos (in  the hidden sector). 
These authors found the following constraint:  $G_{\chi}\lesssim 2.8\times 10^{10} G_{\rm F}$. 
In this study using the current $\rm ^8B$ neutrino flux measured by several solar neutrino detectors, and following a similar strategy to the previous work, we can put an independent constraint on the value of  $G_{\chi}$. Several neutrino experiments~\citep[e.g.,][]{2010PhRvD..82c3006B,2017arXiv170709279A} have reported that the survival probability of electron neutrinos at  $E=8.9 MeV$ is  $P_e(E=8.9 MeV)=0.29\pm 0.1$. This value is consistent with the standard (three active) neutrino  flavor   oscillation model
and the standard solar model. We found that in order to obtain a  $P_e$ that is in agreement with the
$\rm ^8B$ neutrino data, we must have $G_\chi\lesssim 0.5 \, 10^9 G_{\rm F}$.
 
\smallskip 
In conclusion, we have shown that if the neutrino spectroscopic measurements in the near future
 can be used to test new particle physics models, even possible interactions between sterile neutrinos and dark matter, the $\rm ^8B$ and $\rm hep$ electron neutrino spectra will the best ones to use to look for this new type of phenomenon.

\begin{acknowledgments}
The author is grateful to the anonymous referee for the valuable comments and suggestions that improved significantly the contents and presentation of the article.
The author thanks the Funda\c c\~ao para a Ci\^encia e Tecnologia (FCT), Portugal, for the financial support
to the Center for Astrophysics and Gravitation (CENTRA/IST/ULisboa) 
 through the Grant No. UID/FIS/00099/2013.
\end{acknowledgments}

\bibliographystyle{yahapj}

\end{document}